\documentstyle[12pt,epsf]{article}
\makeatletter
\@addtoreset{equation}{section}
\makeatother

\title{New hydrogen-like potentials\footnote{Published in Lett. Math. 
Phys. {\bf 8}, 337-343 (1984)}}

\bigskip

\author{David J.~Fern\'andez C. \\
        Departamento de F\'\i sica, CINVESTAV-IPN \\
        A.P. 14-740, 07000 M\'exico D.F., MEXICO \\
        email: david@fis.cinvestav.mx}

\bigskip

\date{}

\begin{document}
\maketitle
\thispagestyle{empty}
\bigskip

\begin{abstract}
{\small Using the modified factorization method employed in \cite{mi84},
we construct a new class of radial potentials whose spectrum for $l=0$
coincides exactly with that of the hydrogen atom. A limiting case of our
family coincides with the potentials previously derived by Abraham and
Moses \cite{am80}.}
\end{abstract}

\bigskip\bigskip

\section{Introduction.}

In almost all exactly soluble spectral problems of the energy operator,
the possibility of obtaining an exact expression for the energy levels and
the corresponding eigenstates stems from the group theoretical symmetries
which lead to the algebraic factorization method \cite{ih51}-\cite{mo69}. 
This method, until quite recently, appeared to have been fully explored. 
Yet, from time to time, some new potential arises for which the
factorization method can be applied in a new way \cite{bw83}. Recently, a
variant of the factorization method proposed by Mielnik \cite{mi84} has
permitted a description of a new class of potentials whose spectra are
identical to those of the harmonic oscillator \cite{am80}-\cite{ni81}. 
Below we apply this method to show that there is a one-parameter class of
radial potentials, different from a Coulomb field for which the sequence
of energy eigenvalues for $l=0$ is exactly as for the hydrogen atom. 

\section{Hydrogen potential: standard and generalized factorizations.}

As is well known, the hydrogen atom eigenproblem:
\begin{equation}
\left[- \frac{\hbar^2}{2m}\nabla^2 - \frac{e^2}{r}
\right]\psi({\bf r}) = E \psi({\bf r}),
\end{equation}
which, after the separation of the angular variables $\psi({\bf r}) =
Y(\theta,\varphi)R(r)$ leads to a sequence of radial problems: 
\begin{equation}
\frac{1}{r}\left[ - \frac{d^2}{dr^2} + \frac{l(l+1)}{r^2} - \frac{2}{r}
\right]rR = \lambda R,
\end{equation}
where $l=0,1,2,\dots$ labels the angular momentum eigenvalues, $r$ is a
new dimensionless radial coordinate, and the functions $R = \{R(r)\} \
(0\leq r<+\infty)$ form a Hilbert space ${\cal H}_r$ with the scalar
product defined by: $(R,R') = 4\pi\int_0^{+\infty}\overline{R(r)}R'(r) r^2
dr$. It is also well known that the radial Hamiltonians
\begin{equation}
H_l = \frac{1}{r}\left[ - \frac{d^2}{dr^2} + \frac{l(l+1)}{r^2} -
\frac{2}{r}\right]r
\end{equation}
admit a sequence of `factorized forms' which allows us to find very simply
the energy levels \cite{ih51}. These forms are
\begin{eqnarray}
&& H_l = a_l^+a_l - \frac{1}{l^2} \label{ihfactor1} \\
&& H_l = a_{l+1} a_{l+1}^+ - \frac{1}{(l+1)^2} \label{ihfactor2}
\end{eqnarray}
where
\begin{eqnarray}
&& a_l = \frac{1}{r}\left[\frac{d}{dr} + \frac{l}{r} - \frac{1}{l}\right]
r, \label{iha}\\ && a_l^+ = \frac{1}{r}\left[-\frac{d}{dr} +
\frac{l}{r} - \frac{1}{l}\right]r. \label{iha+}
\end{eqnarray}
Now, following the method in \cite{mi84}, we would like to ask two
questions. Is this representation unique? Are $a_l$ and $a_l^+$ the only
first-order differential operators for which
(\ref{ihfactor1})-(\ref{ihfactor2})  hold? It turns out that the answer is
negative. Put:
\begin{eqnarray}
&& A_l =\frac{1}{r}\left[\frac{d}{dr} + \beta_l(r)\right]r \label{mia} \\
&& A_l^+ =\frac{1}{r}\left[-\frac{d}{dr} + \beta_l(r)\right]r \label{mia+}
\end{eqnarray}
and demand that the formula analogous to (\ref{ihfactor1}) should be again
valid:
\begin{equation}
A_l^+ A_l = H_l + \frac{1}{l^2}. \label{mifactor}
\end{equation}
This leads to
\begin{equation}
-\beta_l'(r) + \beta_l^2(r) = \frac{l(l+1)}{r^2} - \frac{2}{r} +
\frac{1}{l^2}. \label{riccati}
\end{equation}
Taking into account that we already have one particular solution, $\beta_l
= (l/r) - (1/l)$ the general solution of (\ref{riccati}) can be easily
obtained. Denote
\begin{equation}
\beta_l = \frac{l}{r} - \frac{1}{l} + \phi_l(r).
\end{equation}
Then
\begin{equation}
-\phi_l' + \frac{2l}{r} \phi_l - \frac{2}{l} \phi_l + \phi_l^2 = 0.
\end{equation}
After introducing a new function, $X_l = 1/\phi_l$, we obtain:
\begin{equation}
X_l' + \frac{2l}{r}X_l - \frac{2}{l}X_l + 1 = 0
\end{equation}
with the general solution
\begin{equation}
X_l = \left(\gamma_l - \int_0^r y^{2l}e^{-2y/l}dy\right)r^{-2l} e^{2r/l};
\quad \gamma_l\in R
\end{equation}
and, therefore,
\begin{equation}
\beta_l(r) = \frac{l}{r} - \frac{1}{l} + \frac{r^{2l}e^{-2r/l}}{\gamma_l
- \int_0^r y^{2l}e^{-2y/l}dy}. \label{solgral}
\end{equation}
The commutator of the new operators $A_l$ and $A_l^+$ is not a number:
\begin{equation}
[A_l, A_l^+] = 2\beta_l'(r).
\end{equation}
Thus, we can apply the factorization method in a new way.

Taking the product $A_lA_l^+$, instead of $A_l^+A_l$ in (\ref{mifactor}),
one has
\begin{equation}
A_lA_l^+ = A_l^+A_l + [A_l, A_l^+] = H_l + \frac{1}{l^2} + 2\beta_l'(r) =
\tilde H_{l-1} + \frac{1}{l^2}
\end{equation}
where $ \tilde H_{l-1}$ is a new Hamiltonian:
\begin{equation}
\tilde H_{l-1} = H_l + 2\beta_l'(r) = \frac{1}{r} \left[-\frac{d^2}{dr^2}
+ \tilde V_{l-1}(r)\right] r, 
\end{equation}
with
\begin{equation}
\tilde V_{l-1}(r) = -\frac{2}{r} + \frac{l(l-1)}{r} + \frac{d}{dr} \left[ 
\frac{2r^{2l}e^{-2r/l}}{\gamma_l
- \int_0^r y^{2l}e^{-2y/l}dy}\right], \ l=1,2,3,\dots
\end{equation}
If
$$
\gamma_l > (2l)!\left(\frac{l}{2}\right)^{2l+1}
$$
or $\gamma_l<0$ for a fixed $l$, the third term has no singularities. 
Furthermore, $\tilde V_{l-1}(r) \rightarrow 0$ as $r\rightarrow \infty$,
and so we obtain a one-parameter family of new self-adjoint Hamiltonians. 
To find the spectra of $\tilde H_{l-1}$ note that: 
\begin{equation}
\tilde H_{l-1} A_l = \left(A_l A_l^+ - \frac{1}{l^2}\right) A_l = A_l
\left( A_l^+ A_l - \frac{1}{l^2}\right) = A_l H_l.
\end{equation}
This implies that if $\{R_{nl}\}$ are eigenvectors of $H_l$ with
eigenvalues $\lambda_n$, $\{A_lR_{nl}\}$ are eigenfunctions of $\tilde
H_{l-1}$ with the same eigenvalues. Furthermore
\begin{equation}
A_lR_{nl} = a_lR_{nl} + \frac{r^{2l}e^{-2r/l}}{\gamma_l
- \int_0^r y^{2l}e^{-2y/l}dy}R_{nl}
\end{equation}
are square-integrable functions. They are orthogonal due to
\begin{equation}
(A_lR_{nl},A_lR_{n'l}) = (A_l^+ A_l R_{nl}, R_{n'l}) = \left(\lambda_n +
\frac{1}{l^2}\right)\delta_{nn'}.
\end{equation}

As one can also check, the operator $A_l$ maps the continuous spectrum
subspace of $H_l$ into the corresponding continuous spectrum space of
$\tilde H_{l-1}$. However, the operator $A_l$ does not map the Hilbert
space ${\cal H}_r$ of radial wave functions into the whole of ${\cal
H}_r$. What remains to be examined, similarly like in \cite{mi84}, is the
`missing vector' $\tilde R_{l \ l-1}$ orthogonal to all vectors of form
$A_lR (R\in{\cal H}_r)$: 
\begin{equation}
(\tilde R_{l\ l-1},A_lR) = (A_l^+\tilde R_{l\ l-1},R)\matrix{ \cr
\equiv \cr {}^R} 0.
\end{equation}
This means that $\tilde R_{l\ l-1}$ is obtained form the first-order
differential equation
\begin{equation}
A_l^+\tilde R_{l\ l-1} = \frac{1}{r}\left[-\frac{d}{dr} +
\beta_l(r)\right]r\tilde R_{l\ l-1} = 0
\end{equation}
whose solution is
\begin{equation}
\tilde R_{l\ l-1} = \frac{c_lr^{l-1}e^{-r/l}}{\gamma_l
- \int_0^r y^{2l}e^{-2y/l}dy}. \label{missing}
\end{equation}
One immediately checks that $\tilde R_{l\ l-1}$ is an eigenvector of
$\tilde H_{l-1}$ with the eigenvalue $-(1/l^2)$: 
\begin{equation}
\tilde H_{l-1} \tilde R_{l\ l-1} = 
\left(A_l A_l^+ - \frac{1}{l^2}\right)\tilde R_{l\ l-1} =
-\frac{1}{l^2}\tilde R_{l\ l-1}.
\end{equation}
For any
$$
\gamma_l > (2l)!\left(\frac{l}{2}\right)^{2l+1} \quad {\rm or}
\quad \gamma_l<0,
$$
equation (\ref{missing}) is a square-integrable function defining the new
ground state of the modified potential. Hence, one can see that when
$$
\gamma_l > (2l)!\left(\frac{l}{2}\right)^{2l+1} \quad {\rm or} \quad
\gamma_l<0,
$$
$\tilde H_{l-1}$ is a new family of self-adjoint radial Hamiltonians with
the same discrete spectrum as $H_{l-1}$. 

The limiting case $\gamma_l = (2l)!(l/2)^{2l+1}$ is worth attention. When
$\gamma_l = (2l)!(l/2)^{2l+1}$ the third term in (\ref{solgral}) tends to
a constant when $r$ tends to $\infty$. Hence, $A_lR_{nl}$ is remaining
square-integrable and $\tilde V_{l-1}(r)$ tends to zero when $r$ tends to
$\infty$. However, the function defining the new ground state is no longer
square-integrable.  Therefore, in this case we obtain a new potential in
which the lowest energy level is missing comparing with $H_{l-1}$. 

\section{The case $l=1$.}

By taking $l=1$ (i.e., starting our procedure from the conventional
Coulomb potential with the simplest centrifugal term) we pass to new
radial functions different from the Coulomb potential but for which the
sequence of energy levels corresponding to the vanishing angular momentum
is exactly the same as for the Hydrogen atom. Our new potentials still
depend of one arbitrary parameter: 
\begin{equation}
\tilde V_0(r) = -\frac{2}{r} + \frac{d}{dr}\left[
\frac{2r^2e^{-2r}}{\gamma_1 - 1/4 + (r^2/2 + r/2 + 1/4)e^{-2r}}\right].
\label{nuevoV0}
\end{equation}
It is interesting to notice that for $\gamma_1 > 1/4$ the potentials
(\ref{nuevoV0}) have the same singularity at $r=0$ and the same asymptotic
behaviour at $r\rightarrow\infty$ as the Coulomb field (see Figure 1). The
corresponding ground states are plotted on Figure 2. 

\begin{figure}[htbp]
\begin{minipage}{6truecm}
\hspace*{3truecm}
\epsfxsize=12truecm
\hskip-1cm\epsfbox{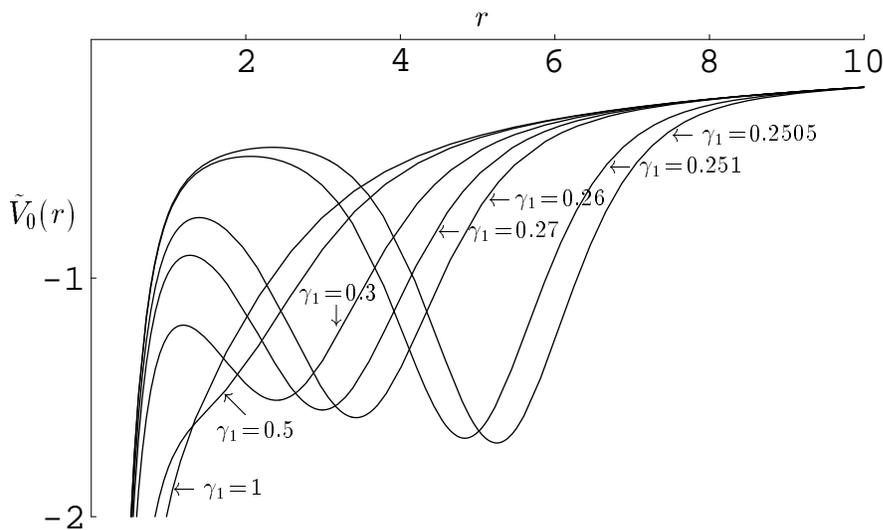}
\end{minipage}
\bigskip
{\caption{\small The potentials $\tilde V_0(r)$ for various values of
$\gamma_1$.}}
\end{figure}

\begin{figure}[htbp]
\begin{minipage}{6truecm}
\hspace*{3truecm}
\epsfxsize=12truecm
\hskip-1cm\epsfbox{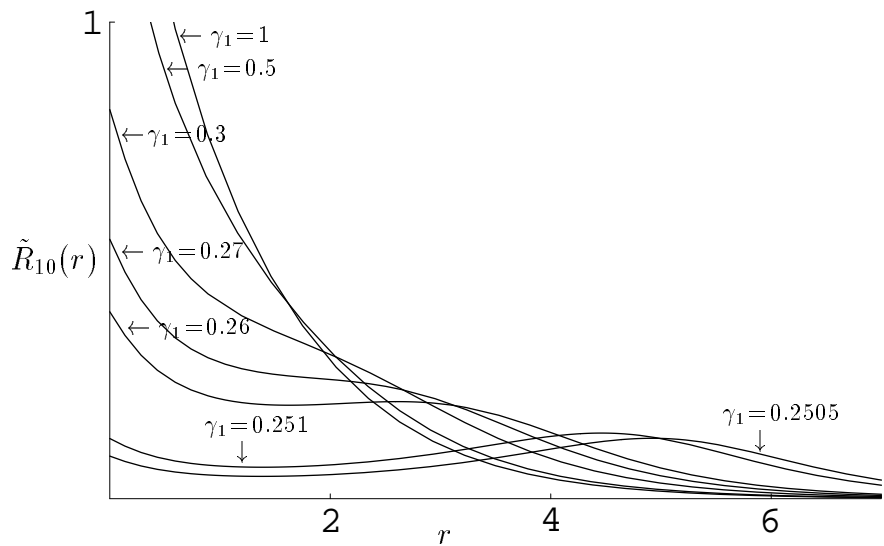}
\end{minipage}
\bigskip
{\caption{\small The ground states $\tilde R_{10}(r)$ for various values
of $\gamma_1$.}}
\end{figure}

The critical value of $\gamma_1\rightarrow 1/4$ is of interest. Our
potential (\ref{nuevoV0}) then tends to: 
\begin{equation}
\tilde V_0(r) = -\frac{2}{r} + 
\frac{16r(r+1)}{(2r^2+2r+1)^2}.
\label{nuevoV0l}
\end{equation}
All the eigenvectors for $\gamma_1\rightarrow 1/4$ converge to some
square-integrable function, except the first one (\ref{missing}) (missing
vector) which then becomes non-normalizable. Therefore, the potential
(\ref{nuevoV0l}) is missing the lowest energy level comparing to the
hydrogen atom. As immediately seen, this case reproduces the one
previously found by Abraham and Moses by applying the Gelfand-Levitan
method (see \cite{am80}). For $\gamma\neq 1/4$, the radial potentials
(\ref{nuevoV0}), as far as we know, have not been known before.

\newpage

\noindent{\bf Acknowledgements.}

\smallskip

I am very much indebted to Dr. Bogdan Mielnik for lending me the
manuscript of his work. Thanks are also due to all my Colleagues at the
Departamento de F\'{\i}sica del CINVESTAV for their interest in this work
and stimulating discussions.


\begin{thebibliography}{99}

\bibitem{ih51} L. Infeld and T.E. Hull, Rev. Mod. Phys. {\bf 23}, 21
(1951).

\bibitem{pleb} J. Pleba\~nski, {\it Notes from lectures on elementary
quantum mechanics}, CINVESTAV, M\'exico (1966). 

\bibitem{mo69} M. Moshinsky, {\it The harmonic oscillator in modern
physics}, in {\it From atoms to quarks}, Gordon and Breach, New York
(1969). 

\bibitem{bw83} D. Basu and K.B. Wolf, J. Math. Phys. {\bf 24}, 478 (1983). 

\bibitem{mi84} B. Mielnik, J. Math. Phys. {\bf 25}, 3387 (1984). 

\bibitem{am80} P.B. Abraham and H.E. Moses, Phys. Rev. A {\bf 22}, 1333
(1980). 

\bibitem{ng81} M.M. Nieto and V.P. Gutschick, Phys. Rev. D {\bf 23}, 922
(1981). 

\bibitem{ni81} M.M. Nieto, Phys. Rev. D {\bf 24}, 1030 (1981). 

\end{thebibliography}
\end{document}